\documentclass[aps,prl,twocolumn,showpacs,superscriptaddress]{revtex4}%
\usepackage{color}
\usepackage{amsfonts}
\usepackage{amsmath}
\usepackage{amssymb}
\usepackage{graphicx}
\usepackage{float}
\usepackage{bm}%

\begin{document}
\title{ Uniaxial-strain detwinning of CaFe$_{2}$As$_{2}$ and BaFe$_{2}$As$_{2}$:
optical and transport study }
\author{M.~A.~Tanatar}
\email[Corresponding author: ]{tanatar@ameslab.gov}
\affiliation{Ames Laboratory, Ames, Iowa 50011, USA}
\author{E.~C.~Blomberg}
\affiliation{Ames Laboratory, Ames, Iowa 50011, USA}
\affiliation{Department of Physics and Astronomy, Iowa State University, Ames, Iowa 50011,
USA }
\author{A.~Kreyssig}
\affiliation{Ames Laboratory, Ames, Iowa 50011, USA}
\author{M.~G.~Kim}
\affiliation{Ames Laboratory, Ames, Iowa 50011, USA}
\affiliation{Department of Physics and Astronomy, Iowa State University, Ames, Iowa 50011,
USA }

\author{N.~Ni}
\affiliation{Ames Laboratory, Ames, Iowa 50011, USA}
\affiliation{Department of Physics and Astronomy, Iowa State University, Ames, Iowa 50011,
USA }
\author{A.~Thaler}
\affiliation{Ames Laboratory, Ames, Iowa 50011, USA}
\affiliation{Department of Physics and Astronomy, Iowa State University, Ames, Iowa 50011,
USA }

\author{S.~L.~Bud'ko}
\affiliation{Ames Laboratory, Ames, Iowa 50011, USA}
\affiliation{Department of Physics and Astronomy, Iowa State University, Ames, Iowa 50011,
USA }
\author{P.~C.~Canfield}
\affiliation{Ames Laboratory, Ames, Iowa 50011, USA}
\affiliation{Department of Physics and Astronomy, Iowa State University, Ames, Iowa 50011,
USA }
\author{A.~I.~Goldman}
\affiliation{Ames Laboratory, Ames, Iowa 50011, USA}
\affiliation{Department of Physics and Astronomy, Iowa State University, Ames, Iowa 50011,
USA }
\author{I.~I.~Mazin}
\affiliation{Naval Research Laboratory, code 6390, Washington, D.C. 20375, USA}
\author{R.~Prozorov}
\affiliation{Ames Laboratory, Ames, Iowa 50011, USA}
\affiliation{Department of Physics and Astronomy, Iowa State University, Ames, Iowa 50011,
USA }
\date{8 March 2010}

\begin{abstract}
The parent compounds of iron-arsenide superconductors, $A$Fe$_{2}$As$_{2}$
($A$=Ca, Sr, Ba), undergo a tetragonal to orthorhombic structural transition
at a temperature $T_{\mathrm{TO}}$ in the range 135 to 205~K depending on the
alkaline earth element. Below $T_{\mathrm{TO}}$ the free standing crystals
split into equally populated structural domains, which mask intrinsic,
in-plane, anisotropic properties of the materials. Here we demonstrate a way
of mechanically detwinning CaFe$_{2}$As$_{2}$ and BaFe$_{2}$As$_{2}$. The
detwinning is nearly complete, as demonstrated by polarized light imaging and
synchrotron $X$-ray measurements, and reversible, with twin pattern restored
after strain release. Electrical resistivity measurements in the twinned and
detwinned states show that resistivity, $\rho$, decreases along the
orthorhombic $a_{o}$-axis but increases along the orthorhombic $b_{o}$-axis in
both compounds. Immediately below $T_{\mathrm{TO}}$ the ratio $\rho_{bo}/
\rho_{ao}$ = 1.2 and 1.5 for Ca and Ba compounds, respectively. Contrary to
CaFe$_{2}$As$_{2}$, BaFe$_{2}$As$_{2}$ reveals an anisotropy in the nominally
tetragonal phase, suggesting that either fluctuations play a larger role above
$T_{\mathrm{TO}}$ in BaFe$_{2}$As$_{2}$ than in CaFe$_{2}$As$_{2}$, or that
there is a higher temperature crossover or phase transition.

\end{abstract}

\maketitle




\section{Introduction}

The parent compounds of high-transition temperature, $T_{c}$, iron arsenide
superconductors, $R$FeAsO ($R$ stands for rare earth) \cite{Hosono} and
$A$Fe$_{2}$As$_{2}$ ($A$ = Ca, Sr, Ba, we denote compounds as A122 in the
following) \cite{Rotter,torikachvili} undergo a tetragonal to orthorhombic
structural transition upon cooling below a transition temperature,
$T_{\mathrm{TO}}$, accompanied or followed by an antiferromagnetic ordering at
$T_{\mathrm{M}}$ \cite{AF}. In free standing single crystals, this transition
leads to the formation of twin domains of four types\cite{domains,chinese}. In
Ba(Fe$_{1-x}$Co$_{x}$)$_{2}$As$_{2}$, $T_{\mathrm{TO}}$ decreases with
increasing $x$ and reaches $T_{c}$ for $x\approx$0.063 \cite{NiNiCo,domains2,
Andreas-Nandi}. For a range of dopings twin domains coexist with
superconductivity and strongly affect it \cite{domains2,Kirtley}. While the
structural anisotropy (the ratio of the in-plane lattice parameters, $a/b)$ is
very small, the electronic structure in the AFM phase is highly anisotropic,
as manifested in the first principles calculations
\cite{Analitis,Ca122FS,Ca122FS1,H} and in the experiment \cite{Davis}. In
particular, the Fermi surface lacks even approximate 4-fold symmetry. At the
same time, despite a dramatic anisotropy in the Fermi surface geometry,
calculations predict a rather small transport anisotropy (see discussion
below). Intrinsic strong anisotropy is a key component in a popular
theoretical picture of \textquotedblleft nematic\textquotedblright%
\ antiferromagnetic state (see review Ref.\onlinecite{Mazin-Schmalian}
for a
discussion), which is believed to be incipient even outside the long-range
ordered antiferromagnetic phase, and probably plays an important role for
superconductivity\cite{Mazin-Schmalian,Kivelson,Subir,rafael}.
Yet, experimentally it has not been accessible till now, because of the
twinned domain structure masking the internal anisotropy. Therefore, it is
important to reliably obtain and characterize textured, or even better, single
domain samples of theses materials.

The fact that $T_{\mathrm{TO}}$ is below room temperature, makes the
situation very different from the classical case of the cuprate YBa$_{2}%
$Cu$_{3}$O$_{7-\delta}$, where single orientations of the orthorhombic domains
can be stable during sample handling and over the entire experimental
temperature range studied \cite{ybco}. Therefore different approaches are
needed to reach a single domain state. The first attempt to study intrinsic
in-plane anisotropy in the iron arsenides was utilizing a strong magnetic
field to align domains \cite{fisher1}, similar to the effect used for turning
La$_{2}$CuO$_{4}$ into a single-domain state \cite{Lavrov}. The application of
a magnetic field of 14~T led to a change of the domain patterns in
Ba(Fe$_{1-x}$Co$_{x}$)$_{2}$As$_{2}$ ($x$=0.016 and 0.025) and in sample with
$x$=0.025 lead to an increase of one of the populations by 7~\%, (from 54 to
61\%.) This resulted in a non-sinusoidal angular dependence of the
magnetoresistance with a magnitude of several percent, and, because of a small
population imbalance, was taken as a signature of large in-plane resistivity anisotropy.

Here we report nearly complete reversible mechanical detwinning in the parent
compounds CaFe$_{2}$As$_{2}$ and BaFe$_{2}$As$_{2}$. We used uniaxial strain
to drive a region of the crystal into a single domain state and study the
temperature dependence of its electrical resistivity. In both CaFe$_{2}%
$As$_{2}$ and BaFe$_{2}$As$_{2}$ we find a clear effect of detwinning on the
resistivity below $T_{\mathrm{TO}}$, with a resistivity decrease along the
orthorhombic $a_{o}$-direction and an increase along the $b_{o}$ direction.
Here and below we use the notations where $a_{o}>b_{o},$ so that the
ferromagnetic chains run along $b_{o}$. That is to say, the observed
anisotropy is not only \textit{much smaller }than one may anticipate from a
naive orbital-ordering picture, but also the \textit{sign} of the anisotropy
is opposite to what one could expect: the electrons move easier along the
antiferromagnetic direction than along the ferromagnetic one. Interestingly,
both these results also appear in the first principle calculations.

The ratio $\rho_{bo}/\rho_{ao}$ is maximal right below $T_{\mathrm{TO}}$,
and is larger in Ba122 than in Ca122 (1.5 vs 1.2). Despite the fact that the
magnetic moment and the degree of orthorhombicity are increasing upon cooling,
the transport anisotropy is \textit{decreasing}. We discuss possible
explanation of this anomalous behavior below.

Finally, the resistivity of Ba122, but not of Ca122 shows measurable
anisotropy for at  least 30~K above $T_{\mathrm{TO}}$, reflecting nematic
fluctuations above $T_{\mathrm{TO}}$ \cite{Johannes} in the former compound,
consistent with the fact that the phase transition in Ca122 is strongly 1st
order, and in Ba122 2nd order or close to that.

\section{Methods}

\subsection{Experimental}

White--light, optical images were taken at temperatures down to 5~K using a
polarization microscope \textit{Leica DMLM} equipped with the flow-type $^{4}%
$He cryostat, as described in detail in Ref.~\onlinecite{domains}.
High-resolution static images were recorded. The spatial resolution of the
technique is about 1 $\mu$m. Single crystals of Ca122 were grown from Sn flux
as described elsewhere. \cite{sim_order_122Ca,Goldman} Single crystals of
Ba122 were grown from FeAs flux \cite{NiNiCo}.

As schematically shown in Fig.~\ref{1-detwin}, panels (a) and (b), four types
of domains are formed in the orthorhombic phase due to orientational
degeneracy of the direction of orthorhombic distortion \cite{domains}. Since
domains O1 and O2 (and similarly O3 and O4) share a common plane corresponding
to the tetragonal (100) [or, equivalently (010) plane], their formation does
not require lattice deformation and they easily form pairs. This leads to a
slight deviation of the orthorhombic axes in the pair from perfectly mutually
orthogonal orientation, by an angle $\pm\alpha= \pi/2 -2\arctan{(b/a)}$. For
resistivity anisotropy study, it is important that domains with two orthogonal
orientations of orthorhombic $a_{o}$ axis are pairwise intermixed, averaging
the in-plane anisotropy in the twinned state.

Straining the crystal along the tetragonal [110] direction (that become
orthorhombic $a_{o}$ or $b_{o}$ axes in the orthorhombic phase) removes the
degeneracy and leads to preferential orientation of domains with the $a_{o}$
axis along the strain. Technically, this requires application of strain at
45$^{\mathrm{o}}$ to the natural tetragonal [100] and [010] facets of the
crystal \cite{mechanical-detwinning}. Due to both very clean natural growth
faces and difficulty to form a good cleave surface, in our initial attempt to
detwin samples of CaFe$_{2}$As$_{2}$ we used as-grown samples. Mechanical
strain was applied to the sample through thick silver wires (125 $\mu$m
thick), soldered with tin-based alloy \cite{SUST} to the corners of the
sample. Use of silver wires to transmit the strain allows for a gentler sample
deformation, which is very important for a material as soft as CaFe$_{2}%
$As$_{2}$. It also allows electrical contact to the sample for electrical
resistivity measurements. The strain was applied by mechanically deforming a
brass horseshoe with a stainless screw, Fig.~\ref{1-detwin}(c), at room
temperature. The wires were soldered to the pre-strained horseshoe, so that
both strain and a little stress (due to wire stiffness) could be applied.

The domain population of the BaFe$_{2}$As$_{2}$ sample has been analyzed by
high-energy x-ray diffraction. Entire reciprocal planes were recorded using
the method described in detail in Ref.\onlinecite{Kreyssig07} which has been
successfully applied recently to study the domain structure in pnictides
\cite{domains,domains2}. The absorption length of the high energy
(99.3\,keV) x-rays from the synchrotron source (beamline 6ID-D in the MUCAT
sector at the Advanced Photon Source, Argonne) was about 1.5\,mm. This allowed
full penetration through the roughly 0.3\,mm thick sample mounted with its
\textbf{c} direction parallel to the incident $x$-ray beam, which was reduced
to 0.2x0.2~mm$^{2}$ size by a slit system. Therefore, as result each single
measurement averages over the entire sample \textit{volume} selected by the
beam dimension in the (\textbf{ab}) plane and its projection through the
sample along the \textbf{c} direction. The direct beam was blocked by a beam
stop behind the sample. Two-dimensional scattering patterns were measured by a
MAR345 image-plate positioned 1730\,mm behind the sample. During the
recording, the sample was tilted through two independent angles, $\mu$ and
$\eta$, perpendicular to the incident x-ray beam by 3.2\,deg.

\subsection{Theoretical}

First principles calculations were performed using the standard LAPW method as
implemented in the WIEN2k package \cite{blaha}. The experimental lattice and
internal parameters were used for both compounds. In order to estimate
the transport anisotropy
within the constant relaxation time Boltzmann approximation, we have
calculated the anisotropic plasma frequency, $\omega_{p\alpha}^{2}%
=4\pi\left\langle Nv_{\alpha}^{2}\right\rangle ,$ where $N$ is the density of
states per unit volume, $\alpha=x,y$, (with $x$ along $a_o$ direction),
$v_{\alpha}$ is a projection of the Fermi velocity and the brackets denote
averaging over the Fermi surface. Up to 70000 k-points in the Brillouin
zone have been used to perform this averaging.

To test the sensitivity of the
results, we have performed calculations in both Local Density
approximation (LDA), with smaller magnetic moments and larger Fermi surfaces,
and Generalized Gradient approximation (GGA), with larger magnetic moments and
smaller Fermi surfaces.

\section{Results}

\subsection{Detwinning}

\begin{figure}[tb]
\begin{center}
\includegraphics[width=0.85\linewidth]{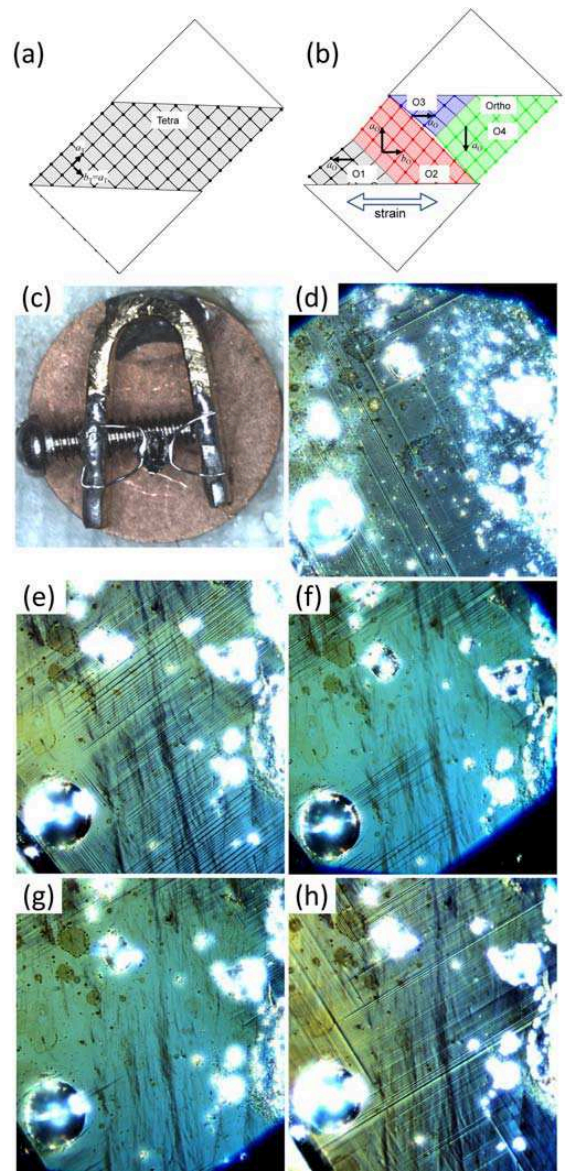}
\end{center}
\caption{(Color online) (a,b) Schematics of domain formation during
tetra-ortho transition. Orthorhombic phase is realized through displacement of
atoms along the diagonal of the tetragonal lattice with doubling the unit cell
volume and 45$^{o}$ rotation of the crystallographic axes \cite{Goldman}. The
domains are distinguished by the direction of the distortion, $a_{o}$, shown for
four domains with the arrows. (c) A single crystal sample of CaFe$_{2}$%
Fe$_{2}$, mounted on a horseshoe. (d-h) Polarized light image of the crystal
without external strain (d), strained to a progressively higher values (e-g),
and after strain release (h). }%
\label{1-detwin}%
\end{figure}

The series of pictures in Fig.~\ref{1-detwin} shows polarized optical images
of the Ca122 sample, taken at 5~K, in the area between the contacts without
external strain (d), at progressively increasing strain, (e) through (g), and
after strain release (h). On the application of strain the domain pattern in
the central area between contacts changes rapidly and eventually a domain free
region is formed. However, a pattern of randomly oriented domains is still
seen away from the central part, reflecting the distribution of strain over
the sample area and its concentration close to the soldered contacts. Strain
release restores the domain pattern, i.e. detwinning is reversible.

\begin{figure}[tb]
\begin{center}
\includegraphics[width=1.0\linewidth]{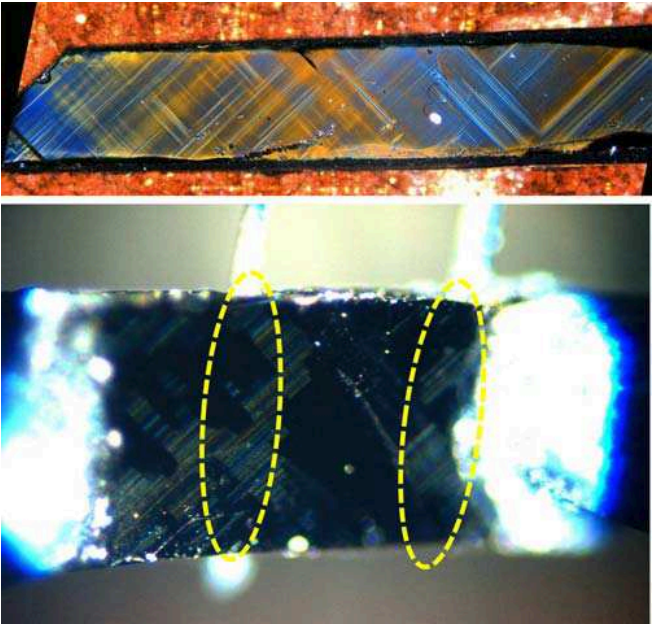}
\end{center}
\caption{(Color online) A polarized light image (5~K, $T \ll T_{\mathrm{TO}}$)
of a single-crystal strip of CaFe$_{2}$As$_{2}$ cut along the tetragonal [110]
direction before contacts were made (top). The sample was mounted in a usual
four probe contact configuration with potential contacts located on the sample
surface opposite to the surface in the image. Strain was applied to the sample
at room temperature through current contacts and the image was taken at 5~K on
a clear surface under the potential contacts. A big single-domain area can be
seen between potential contacts, however, large heavily twinned areas are
found above the contacts (circles in the image), revealing notable local
reduction of strain by surface tension at the contact points. }%
\label{2-contacts}%
\end{figure}

Encouraged by this initial success, we introduced two modifications into the
sample detwinning process. First, samples were cut into strips to allow a
homogeneous distribution of strain between contacts. The image of the sample
after cutting, Fig.~\ref{2-contacts}(a), reveals a pattern of random domains,
as is typical for an unstrained sample. Second, we attached contacts in a
conventional four-probe configuration. The thick wire contacts were made at
the ends of the sample and served as current contacts. The potential leads
were made from a much thinner, and thus less rigid, 50 $\mu$m silver wire long
enough to avoid generation of additional strain. The potential contacts were
mounted on the rear surface of the crystal, so that we could monitor the
detwinning in a polarized optical microscope. Fig.~\ref{2-contacts}(b) shows
an optical image of a sample strained through current contacts. Although there
is a clear change in the domain distribution upon the application of strain,
showing that detwinning affects the whole sample thickness \cite{c-axis}, two
big unstrained areas can be seen in the central region, which corresponds to
the areas above the potential contacts on the rear surface. This observation
clearly shows that the contacts affect strain distribution in the sample and
do not allow homogeneous sample deformation.

\begin{figure}[tb]
\begin{center}
\includegraphics[width=.75\linewidth]{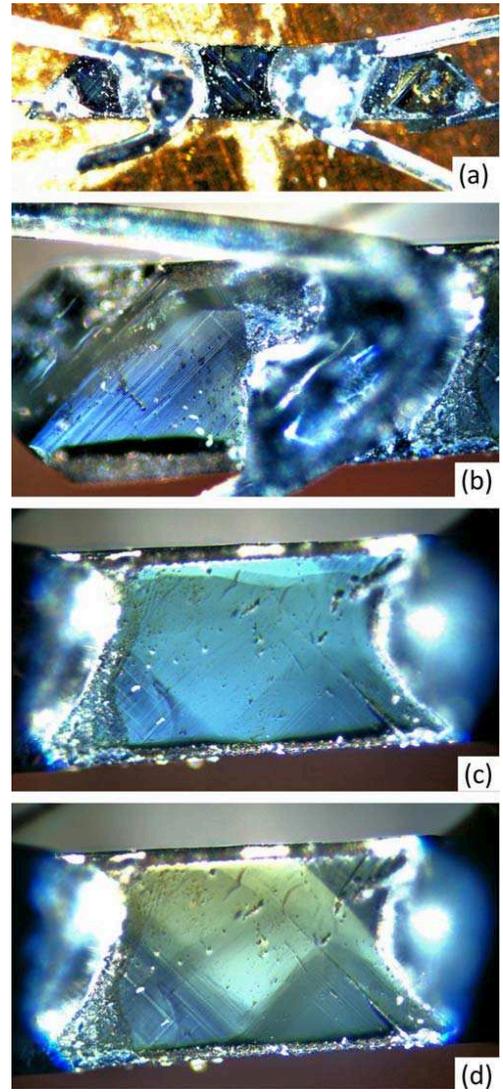}
\end{center}
\caption{(Color online) (a) A polarized light image (5~K, $T \ll
T_{\mathrm{TO}}$) of a single crystalline strip sample of CaFe$_{2}$As$_{2}$
with soldered contacts before being mounted on a horseshoe. On application of
strain in a horizontal direction through the potential contacts, a strain-free
area at the left end of the sample remains heavily twinned (b), while the
central area turns into a nearly single domain state, with small twinned areas
at the bottom of the image (c). After three thermal cycles from the 300~K to
5~K range, strain in the sample is partially released (d), however, large
single-domain areas remain, demonstrating the reproducibility of the
strain-induced detwinning. }%
\label{3-potential-strain}%
\end{figure}

In an attempt to further improve detwinning for resistivity measurements, we
started sample deformation through potential contacts. In this case only a
central part of the sample, between the potential contacts, is deformed.
Simultaneously, the parts of the sample between the thin-wire current contacts
and the thick-wire potential contacts remain strain-free. Imaging of the
sample at 5~K during different stages of the sample preparation and straining
is shown in Fig.~\ref{3-potential-strain}. As can be seen in a series of
images, the central part of the sample is predominantly free of domains, while
a clear pattern of domains can be seen in the unstrained part of the sample.
The pattern of domains disappears abruptly at $T_{\mathrm{TO}}$, the strained
part abruptly changes color at $T_{\mathrm{TO}}$, signaling the disappearance
of the orthorhombic distortion. Formation of the orthorhombic phase is
homogeneous through most of the sample, except for two small patches close to
the contacts.

\subsection{Resistivity of CaFe$_{2}$As$_{2}$}

\begin{figure}[tb]
\begin{center}
\includegraphics[width=1.0\linewidth]{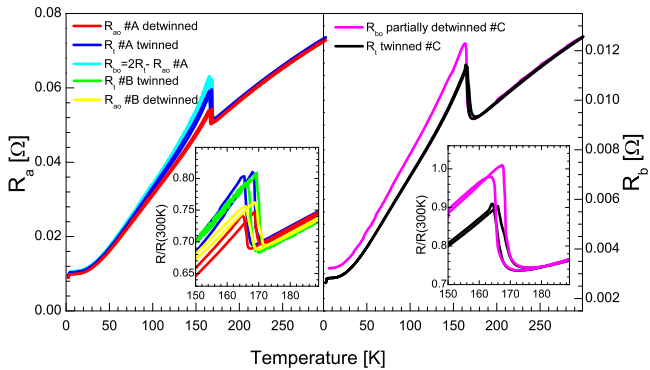}
\end{center}
\caption{(Color online) Left panel. Temperature-dependent electrical
resistance measured on a sample A of CaFe$_{2}$As$_{2}$ in twinned ($R_{t}$,
blue curve) and detwinned ($R_{ao}$, red curve) states. The third curve is
obtained as hypothetical $R_{b_{o}}=2R_{t} -R_{a_{o}}$, impossible to measure
in the same contact arrangement, see text for details. Inset shows zoom of the
transition region for samples A and B, revealing diminishing amplitude of the
resistance jump in $\rho_{a_{o}} (T)$ with identical transition sharpness and
position. Right panel shows resistance of sample C, in which resistivity was
measured along $b$-axis in transverse to strain contact configuration (see
Fig.~\ref{5-transverse}) in stress-free and stressed (partially detwinned)
states. Inset shows zoom of the transition area. This data directly show an
increase of the resistance jump in $\rho_{b_{o}}$. }%
\label{4-resistivity}%
\end{figure}

The left panel of Fig.~\ref{4-resistivity} shows the temperature-dependent
electrical resistance, measured on sample A during two successive thermal runs
(cooling from 300~K to 5~K and warming up to 300~K) in the unstressed (blue
curve) and stressed (red curve) states. It is clear that the strain itself is
not large enough to cause a change in the value of the sample resistance at
room temperature, beyond a small ($\sim$ 1\%) systematic resistance increase on
strain change (both increase and decrease) due to a fatigue. The data above
$T_{\mathrm{TO}}$ stay unchanged, below a clear change in resistance is
observed. In particular, the upward jump in the resistivity below
$T_{\mathrm{TO}}$ is notably diminished in the detwinned state. The data in
the unstrained state compare well with the temperature-dependent resistivity
in the standard resistance measurements along the [100] direction
\cite{torikachvili,sim_order_122Ca,budko,anisotropy2}. The inset in the left
panel shows the same data, normalized by the room temperature value, in
comparison with the data for sample B, measured in the same contact
configuration. The data agree very well in the magnitude of the effect. It can
be clearly seen that the position of the transition is not shifted with
strain, the transition remains sharp, and the width of the hysteresis between
the cooling and warming cycles remains the same. This clearly shows that the
transition itself is insensitive to the applied strain, and, in view of the
strong sensitivity of the transition temperature to uniaxial stress
\cite{budko}, the applied strain is very small and homogeneous.

To check if the effects of thermal cycling are important, we took a polarized
light image of the sample A after a resistivity run in the detwinned state.
The optical image of the sample in a third successive thermal run, following
the optical run of Fig.~\ref{3-potential-strain}(c) and resistivity
measurements in Fig.~\ref{4-resistivity}, is shown in a panel (d) in
Fig.~\ref{3-potential-strain}. As can be seen, the strain is slightly released
after thermal cycling, however most of the sample remains domain free. After
this second imaging, the strain in the sample was released, and
temperature-dependent resistivity measurements were undertaken in a
strain-released state, main panel in the left panel of
Fig.~\ref{4-resistivity}. These revealed a resistivity curve very close to the
strain-free samples, restoring the magnitude of the resistivity up-jump to its
initial value.

Since the contact configuration in Fig.~\ref{3-potential-strain} does not
allow for a measurement of $\rho_{bo}$, we inferred its evolution by assuming that
the resistivity in the twinned state is an average of $\rho_{ao}$ and $\rho_{bo}$.
This is strictly true if the population of the domains is random. The
calculated $R_{bo}(T)\equiv2R_{t}-R_{ao}$ is shown in the left panel. The
comparison of $\rho_{bo}(T)$ and $\rho_{ao}(T)$ gives a ratio of 1.2 at the
transition. With cooling, the ratio decreases to 1.05--1.1 (due to the small
absolute value, the effect of fatigue cannot be neglected, and it is hard to
place a more precise value on the low-temperature anisotropy).

\begin{figure}[tb]
\begin{center}
\includegraphics[width=.6\linewidth]{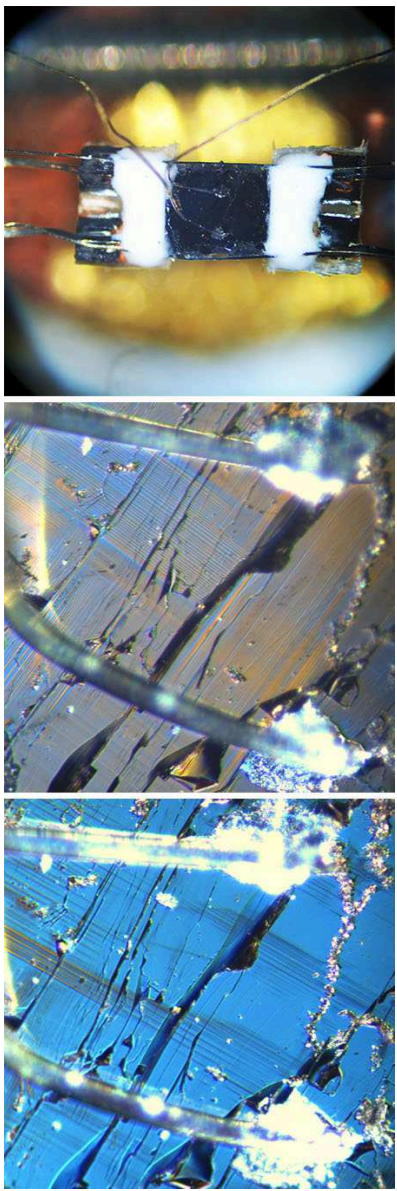}
\end{center}
\caption{(Color online) Top panel. Image of the sample for resistivity
measurements in transverse to the strain (horizontal) direction. The current
contacts are made along the whole length of the top and bottom edges of the
sample, right and left sample edges are covered with with epoxy glue to
provide homogeneous deformation of the sample and wires. Straining force is
applied on horseshoe, similar to Fig.~1c. Potential contacts are made small to
minimize the effect on the strain distribution. Middle and bottom panels show
area close to potential contacts in stress-free and stressed states. }%
\label{5-transverse}%
\end{figure}

To get an independent assessment of the $\rho_{bo}(T)$ we have mounted the sample C
in such a
way that the strain be generated in the direction perpendicular to the
current. This was achieved by soldering the current contacts along the
entire
side of the rectangular sample and straining the sample and the current
contact wires on a horseshoe. The ends of the sample were covered with epoxy
glue and glued to two pieces of fiberglass to provide homogeneous deformation
(see Fig.~\ref{5-transverse}). The area of the potential contacts was
minimized to diminish their effect on the measurements. Sample imaging showed
clear detwinning, see Fig.~\ref{5-transverse}, however not in the entire
sample. In the right panel of Fig.~\ref{4-resistivity} we show the resistivity
obtained in the twinned and partially detwinned states of sample C. The
resistivity indeed shows an increase right below $T_{\mathrm{TO}}$. At the
same time, the temperature-dependent resistance in the tetragonal phase is
unaffected by the strain all the way down to $T_{\mathrm{TO}}$, indicating
that there is no residual nematic ordering above $T_{\mathrm{TO}},$ which is,
in fact consistent with the fact that the phase transition in Ca122 is
strongly 1st order \cite{rafael}. Note that since all measurements are taken
on the same sample with the same contacts, they are free of any possible
geometric uncertainties.

\subsection{Resistivity of BaFe$_{2}$As$_{2}$}

Single crystals of Ba122 were detwinned in the same fashion as Ca122 above,
with optical imaging confirming a homogeneous detwinned state for resistivity
measurements along the $a_{o}$-axis and a predominantly detwinned state for
resistivity measurements along the $b_{o}$-axis. Because of small image
contrast, samples were additionally studied with high-energy synchrotron
$x$-ray as described in the next section.

\begin{figure}[tb]
\begin{center}
\includegraphics[width=1.0\linewidth]{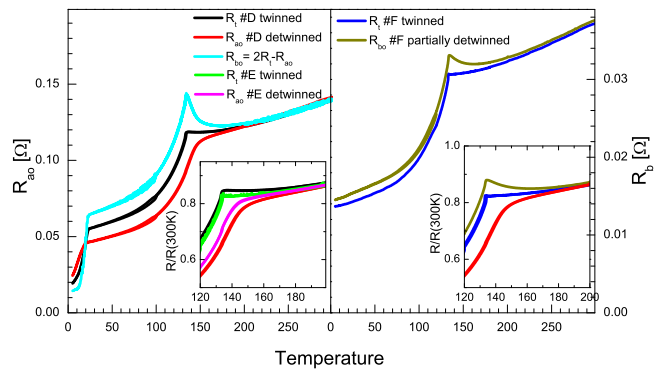}
\end{center}
\caption{(Color online) Left panel. Temperature-dependent electrical
resistance measured on a sample D of BaFe$_{2}$As$_{2}$ in twinned ($R_{t}$,
black curve) and detwinned $R_{ao}$, red curve) states. The third curve is
obtained as hypothetical $R_{bo}=2R_{t}-R_{a_{o}}$, impossible to measure in
the same contact arrangement. Inset shows zoom of the transition region for
samples D and E. Right panel shows resistance for sample F, in which
resistivity was measured along $b$-axis in transverse to strain contact
configuration in stress-free (blue line) and stressed (partially detwinned,
yellow line) states. Inset shows zoom of the transition area in normalized
resistance plot in comparison with resistance of sample D in the detwinned
state. }%
\label{6-resistivityBa}%
\end{figure}

In Fig.~\ref{6-resistivityBa} we show the temperature-dependent electrical
resistivity of Ba122. The left panel shows the resistance measured along the
strain, $R_{ao}$ (sample D), the inset shows the transition area with the data
from sample D as well as from another sample, E. The partial superconductivity
seen in these samples is presumably due to surface strain \cite{Saha}
associated with cleaving and shaping the sample and is not focus of this
study. As for the case of CaFe$_{2}$As$_{2}$, we can infer the resistivity
along the $b_{o}$ axis from $R_{bo} = 2R_{t} -R_{ao}$; the inferred $R_{bo}$
is also shown. The right panel shows resistivity measurements with current
transverse to the strain in the twinned and partially detwinned states on yet
another sample F.

Below $T_{\mathrm{TO}}$ the data from Ba122 is qualitatively similar to that
of Ca122; both show anisotropies right below $T_{\mathrm{TO}},$ of 1.2 (Ca122)
and 1.5 (Ba122). In both materials the anisotropy decreases with cooling below
$T_{\mathrm{TO}},$ that is to say, is anticorrelated with the degree of
orthorhombicity and the long-range magnetic moment. The absolute change of the
anisotropy is similar, 10\% (from 50\% to 40\% in Ba122, from 20\% to 10\% in
Ca122). Note that this similarity holds despite the fact that the residual
resistivity in Ba122 is substantially higher --- the anisotropy of $\rho_{0}$
is also higher. As explained below, this is consistent with the idea that the
transport anisotropy reflects the anisotropy of the carrier velocity.

Another important observation is that in Ca122 the anisotropy drops to 1 right
above the transition, reflecting its 1st order nature, in Ba122 an anisotropy
is detectable at least up to 30 K above the transition, a manifestation of
strong nematic fluctuations above $T_{\mathrm{TO}}$ \cite{Johannes}. This is
consistent with the phase transition in Ba122 being second order or very close
to such, which, in term, was argued to be related to a harder lattice and
weaker magnetoelastic coupling in Ba122.

\subsection{X-ray characterization of detwinned state of BaFe$_{2}$As$_{2}$}

\begin{figure}[tb]
\includegraphics[width=1.0\linewidth]{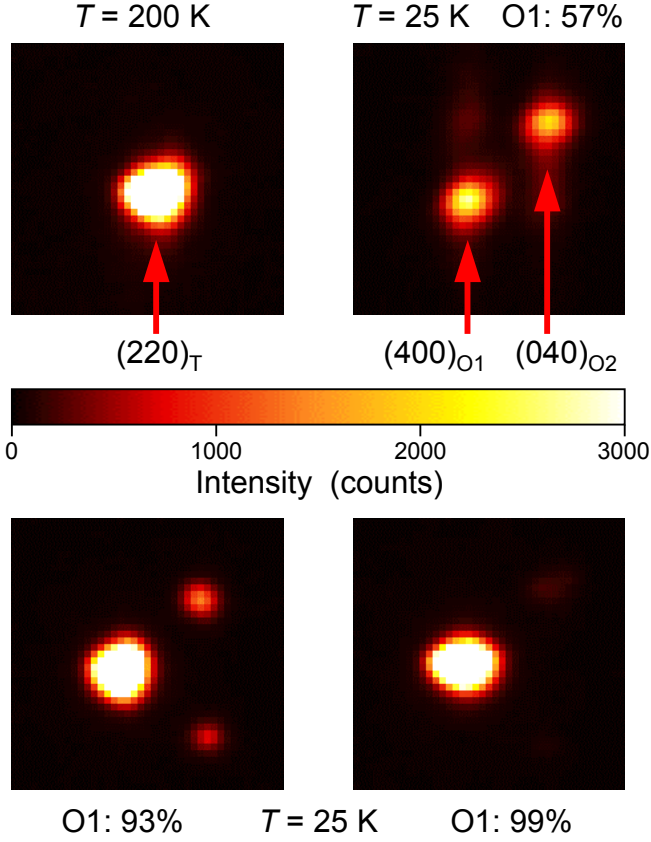}
\caption{Sections of reciprocal ($HK0$) planes around the position of the
tetragonal (220)$_{T}$ reflection recorded from the BaFe$_{2}$As$_{2}$ sample
by high-energy x-ray diffraction as described in the text. In the orthorhombic
state at $T$~= 25\,K, the (400)$_{O1}$ and (040)$_{O2}$ reflections are
related to different domains $O1$ and $O2$, respectively. Analysis of the
intensities in pattern recorded at different positions on the sample, below a
contact in the upper panels and between the contacts in the lower panels, yield
the indicated \%-values for the domain $O1$ population bolstered by the
applied strain.}%
\label{xrayfig1}%
\end{figure}

The upper panels of Fig.~\ref{xrayfig1} show the splitting of the tetragonal
(220)$_{T}$ reflection into the orthorhombic (400)$_{O1}$ and (040)$_{O2}$
reflections produced by domains $O1$ and $O2$, with different orientations,
below the structural transition in a sample area that was nearly strain free.
Illuminating the sample with 200$\mu$m diameter $x$-ray spot at positions
between strain-applying contacts yields the pattern shown in the lower panels
of Fig.~\ref{xrayfig1}. The intensity in the (400)$_{O1}$ reflection is
strongly enhanced indicating that the domains with their orthorhombic
\textbf{a} axis (\textbf{a}\,$>$\,\textbf{b}) along the direction of the
applied strain are more populated. The full penetration of the high-energy
$x$-ray beam through the sample allows a quantitative analysis of the bulk
domain population with an error of approximately 1\,\% by using the relative
intensities in all orthorhombic (400) and (040) reflections.

\begin{figure}[tb]
\includegraphics[width=1.0\linewidth]{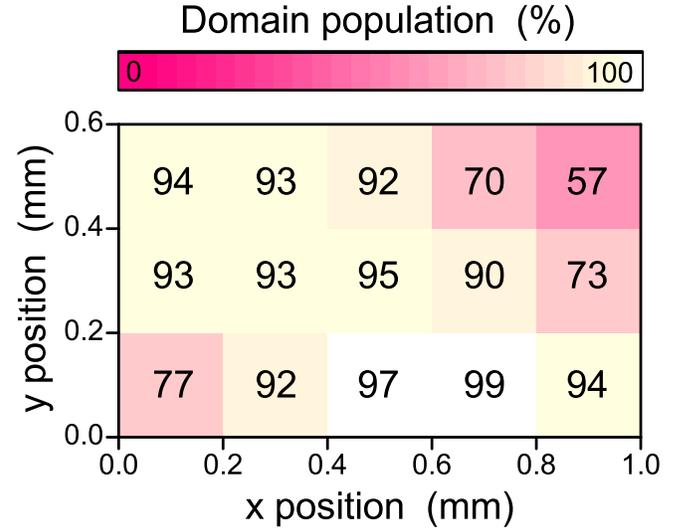}
\caption{Spatial distribution of the population of $O1$ domains in strained
BaFe$_{2}$As$_{2}$. The population was determined from the relative
intensities of the (400)$_{O1}$ and (040)$_{O2}$ peaks, as shown in
Fig.~\ref{xrayfig1}. The area between the contacts, through which the strain
was applied, was probed in 0.2~mm vertical and horizontal steps, determined by
the $x$-ray beam size. The number and color in each pixel correspond to the
percent-value for the domain $O1$ population. The pixel in the upper right
corner corresponds to an area under the contact. }%
\label{xrayfig2}%
\end{figure}

Scanning the x-ray beam across the sample allows a spatially resolved
characterization of the domain population as demonstrated in
Fig.\ref{xrayfig2}. Here we show that the sample is nearly detwinned between
the contacts, whereas the domain population comes closer to 50\% (the value
expected for a random distribution of twins) at positions below the contacts, as shown in the right-top pixel in Fig~\ref{xrayfig2}. This confirms
our conclusion, Fig.~\ref{2-contacts}, that contacts act as anchors for the deformation.

\subsection{Theoretical analysis}

To analyze these results let us write the resistivity in Boltzmann (relaxation
time) approximation, keeping in mind that the relaxation times for holes and
for electrons may be different. Despite the fact that the magnetic structure
of A122 compounds is highly anisotropic, we would assume that the relaxation
times is isotropic, as we will discuss below (in particular, it is important
to keep in mind that the Fermi surfaces, both experimental and calculated, are
much smaller in size that the antiferromagnetic vector). Thus, we write%
\begin{align}
1/\rho &  =1/\rho_{h}+1/\rho_{e}\label{rho}\\
\rho_{i} &  =(4\pi/\omega_{p,i}^{2})(1/\tau_{0,i}+1/\tau_{ph,i}+1/\tau
_{ee,i}+1/\tau_{mag,i}),
\end{align}
where $i$ stands for hole or electrons. The prefactor, $\omega_{p}^{2},$ is
the only quantity directly affected by twinning (except for a possible scattering by
domain walls in $\tau_{0}$ that in principle would go away with detwinning;
however, the twin domain walls are usually weak scatterers). In the classical
Boltzmann theory it is presumed that $\omega_{p}^{2}$ and $\tau_{0}$ are
temperature-independent, which allows one to separate $\rho$ additively into
two terms, a temperature independent residual resistivity, and the temperature
dependent scattering by thermal excitations (phonons, magnons, electrons).
Unfortunately, this is not the case here. The Hall effect studies found that
upon cooling in the AFM phase, the carrier concentration drops drastically, as
reflected by the nearly an order of magnitude reduction in the plasma
frequency, $\omega_{p}^{2},$ and the density of states also decreases
substantially (the latter effect makes $\tau_{0}$ temperature dependent, but
this is not important for us now).

The standard approximation implies an isotropic relaxation time, even when the
bosons providing inelastic scattering are anisotropic. Our first task is to
verify whether this remains a good approximation in our case. Given the large
magnetic anisotropy and large difference in the ferro- and antiferromagnetic
correlation length above $T_{\mathrm{TO}}$ \cite{diallo,diallo1,zhao} one may
think that scattering by the spin fluctuations may be anisotropic as well. To
analyze that, let us recall the classical theory of the inelastic transport
scattering in solids. The scattering rate for the current flowing in a given
direction $i$ is defined by the Fermi golden rule (see, $e.g.,$ Ref.
\onlinecite{Allen}, Eq. 37]):%

\begin{equation}
\frac{1}{\tau_{i}}\approx A\sum_{\mathbf{kk}^{\prime}}v_{\mathbf{k}i}%
^{2}|M_{\mathbf{kk}^{\prime}}|^{2}\delta(\varepsilon_{\mathbf{k}}%
)\delta(\varepsilon_{\mathbf{k}^{\prime}})/\sum_{\mathbf{k}}v_{\mathbf{k}%
i}^{2}\delta(\varepsilon_{\mathbf{k}}),
\end{equation}
where $A$ is an isotropic factor, the denominator is proportional to the
plasma frequency squared, $M_{\mathbf{kk}^{\prime}}$ is the matrix element of
the electron-magnon scattering by a magnon with the wave vector $\mathbf{q=k-k}%
^{\prime}$ and energy $\hbar\omega(\mathbf{q)}$, and the cross-terms involving
$v_{k}v_{k^{\prime}}$ were neglected. The standard argument goes like this:
even though the bosons (magnons, in our case), may be much softer, and more
easily excited by temperature, in one $q$-direction than in the other, the
fact that the scattering processes are integrated twice over the Fermi surface
effectively averages out any spectral anisotropy. Moreover, in fact direct
neutron measurements \cite{diallo1,zhao} show that the magnon dispersions
along $x$ and along $y$ differ by less than 15\%. Furthermore, small size of
the Fermi surfaces in the magnetic state (note that the Hall concentration
drops quite rapidly below $T_{\mathrm{TO}},$ indicating that the Fermi
surfaces start to shrink immediately below the transition \cite{Mazin}) only
lets the magnons with small \textbf{q} participate in scattering, and at small
\textbf{q}s the anisotropy of the magnon spectra is particularly low. Thus a
strong anisotropy in $\tau$ is unlikely, although cannot be excluded.

\begin{figure}[tb]
\includegraphics[height=3.0cm,width = 4 cm]{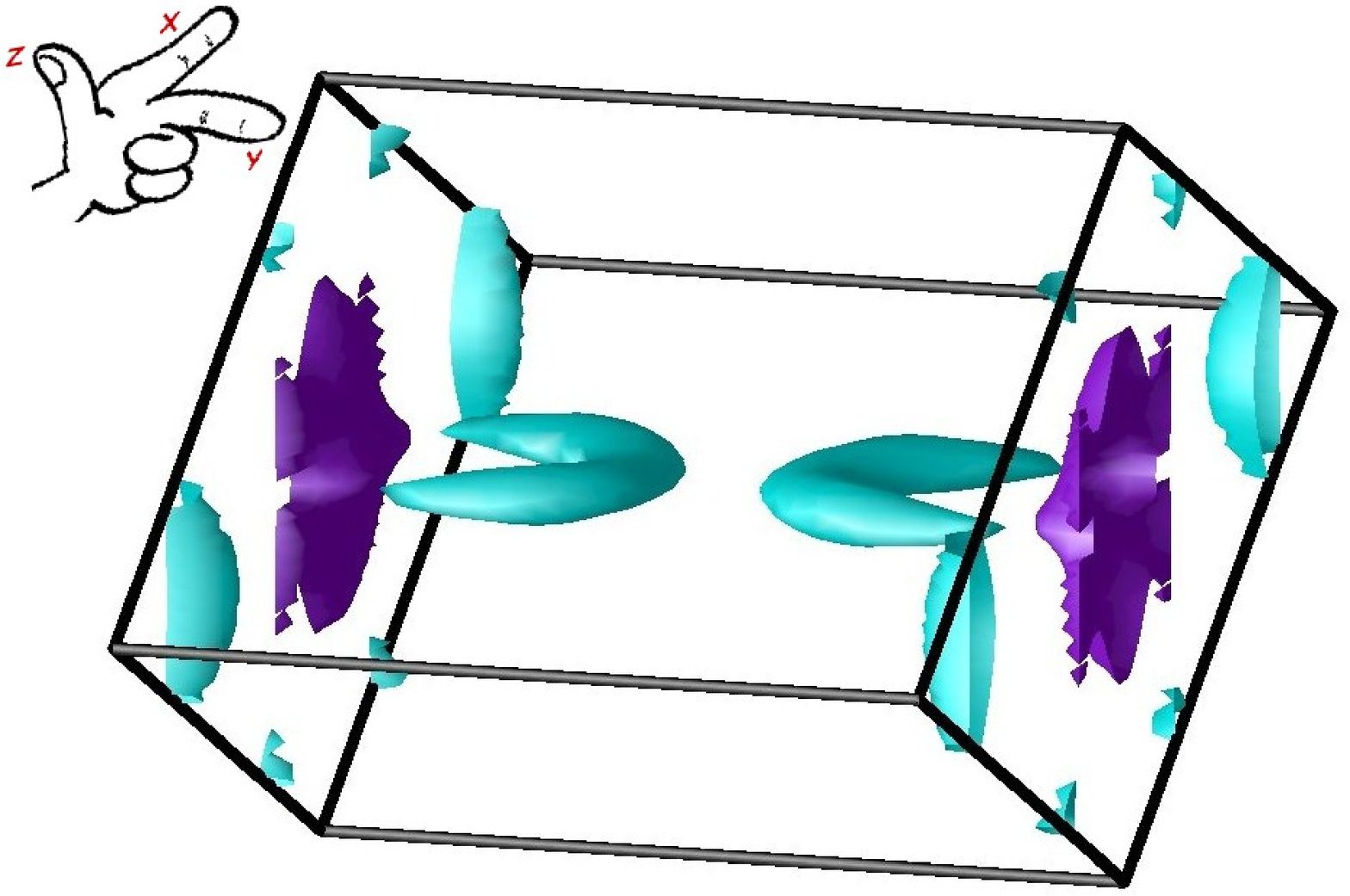}
\hspace{0.05cm}
\includegraphics[height=3.2cm,width = 4 cm]{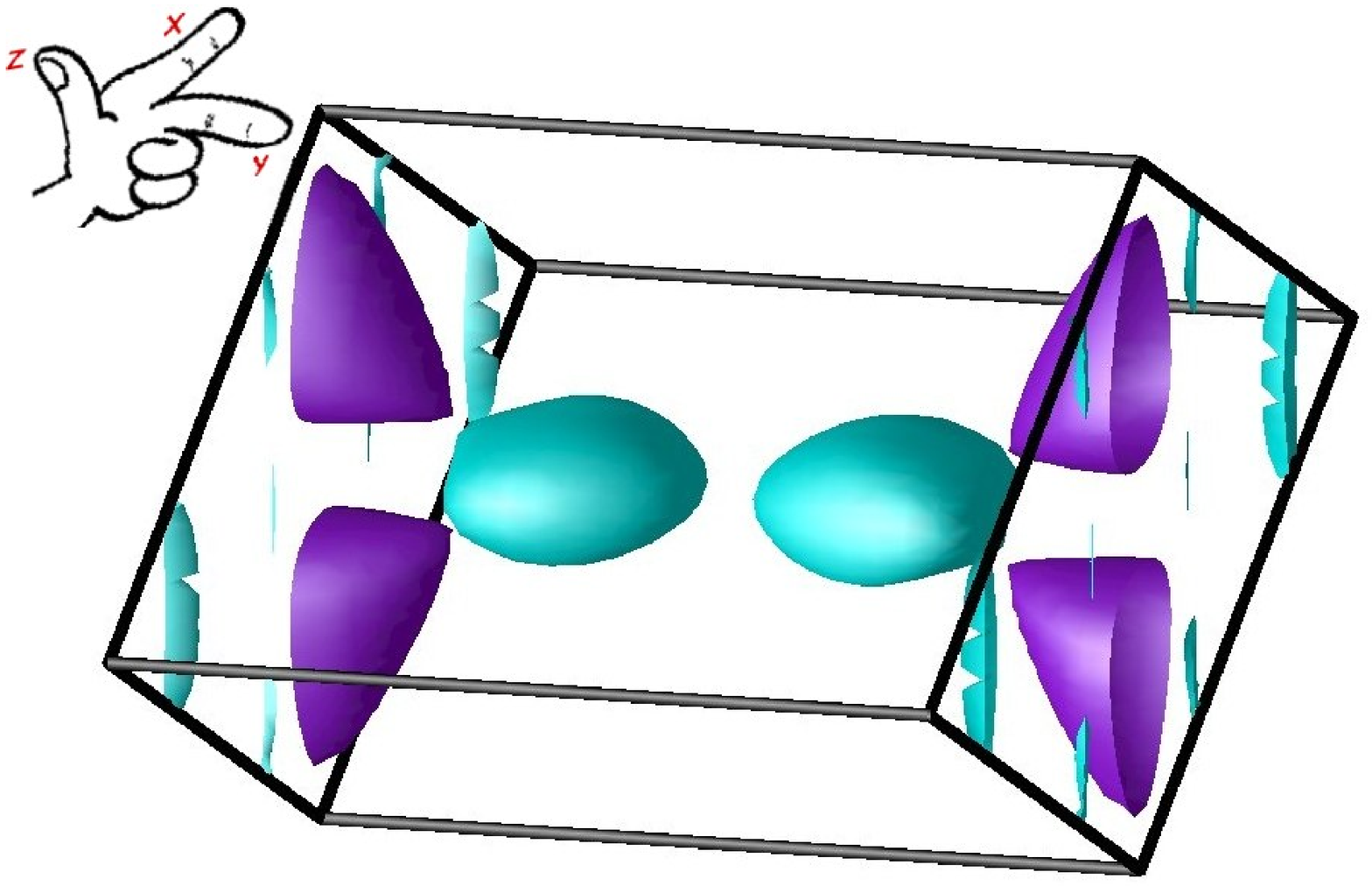}
\vspace{0.2cm}
\includegraphics[width = .45\linewidth]{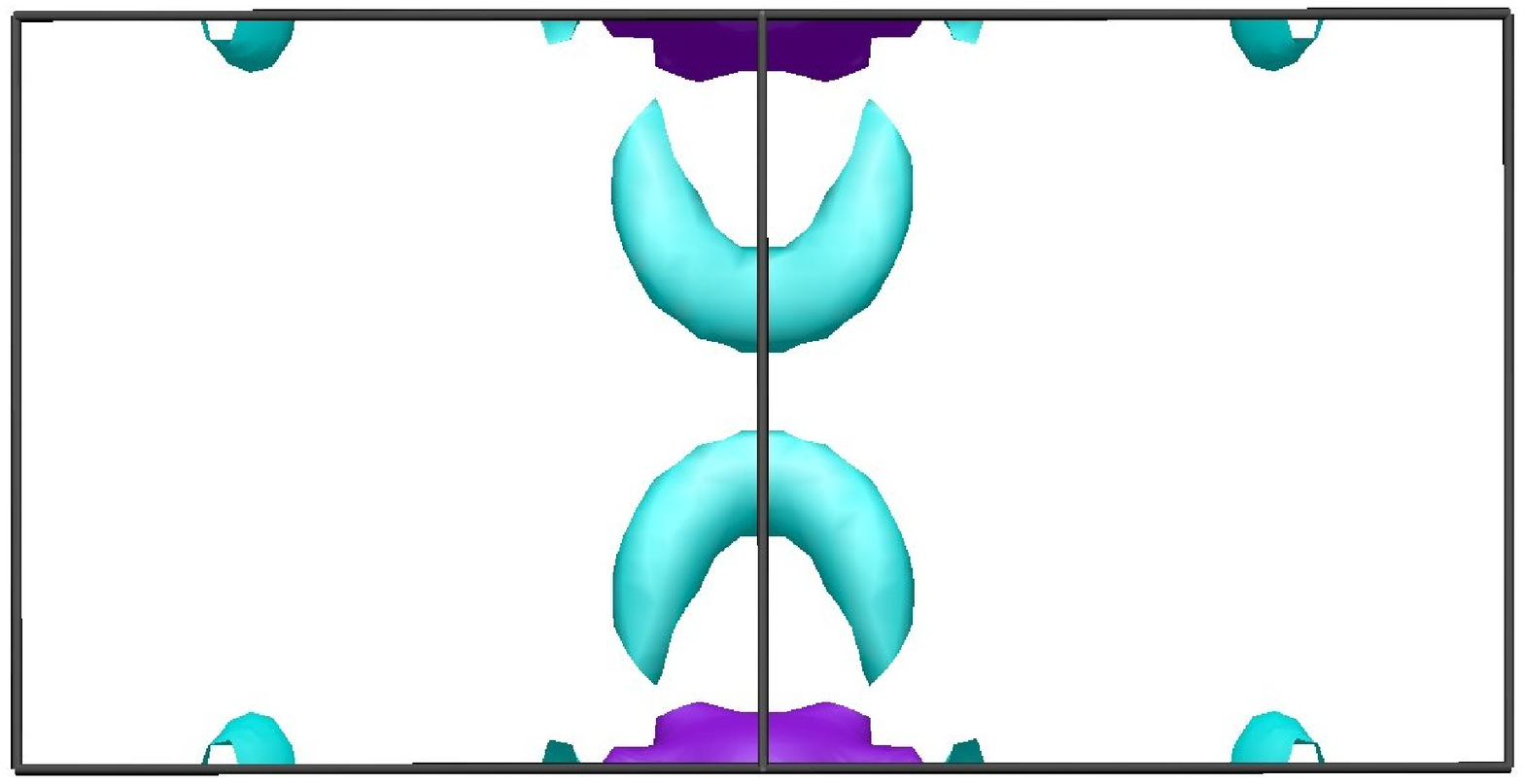}
\hspace{0.05cm}
\includegraphics[width = .45\linewidth]{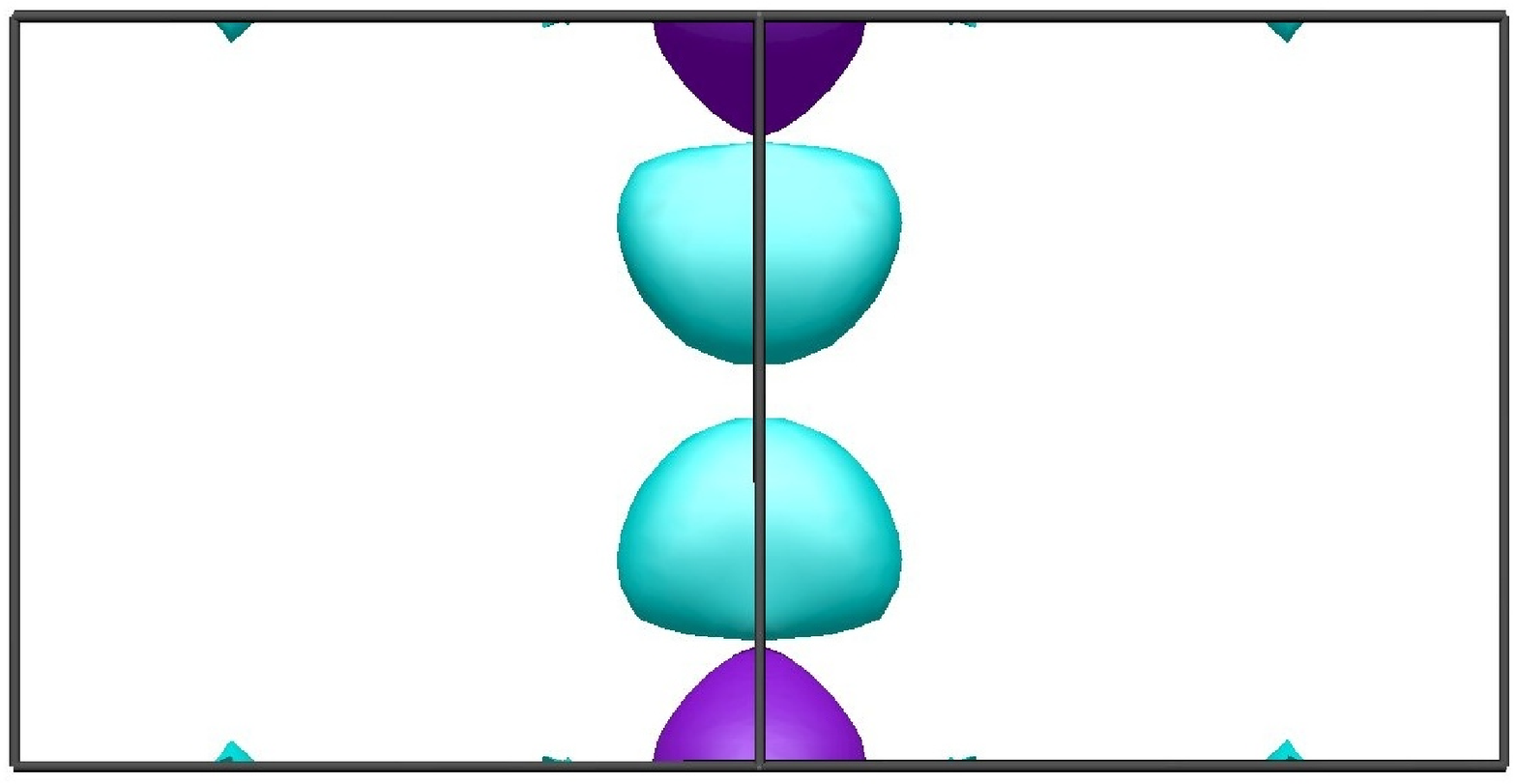}
\vspace{0.1cm}
\includegraphics[width = .45\linewidth]{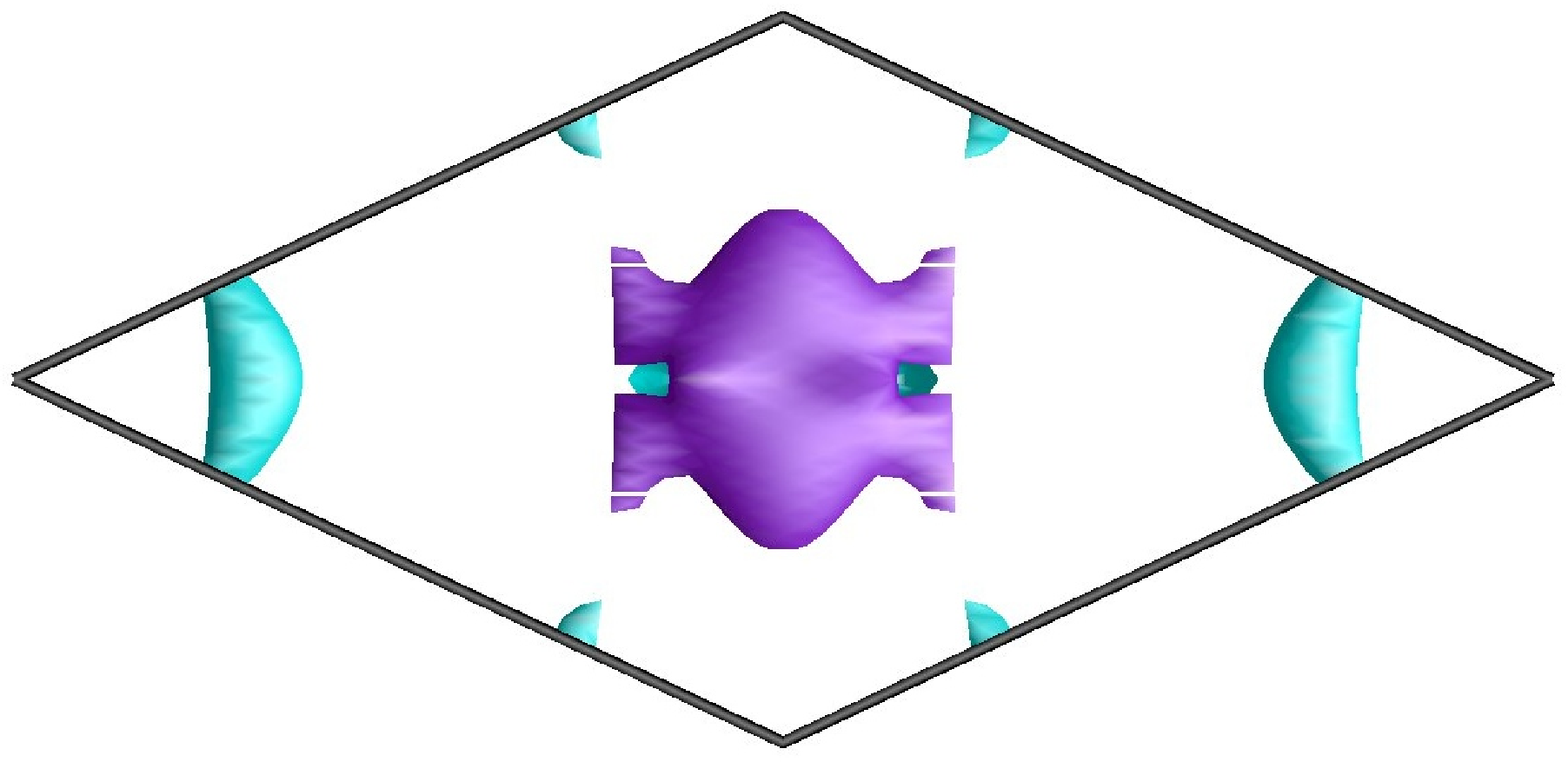}
\hspace{0.05cm}
\includegraphics[width = .45\linewidth]{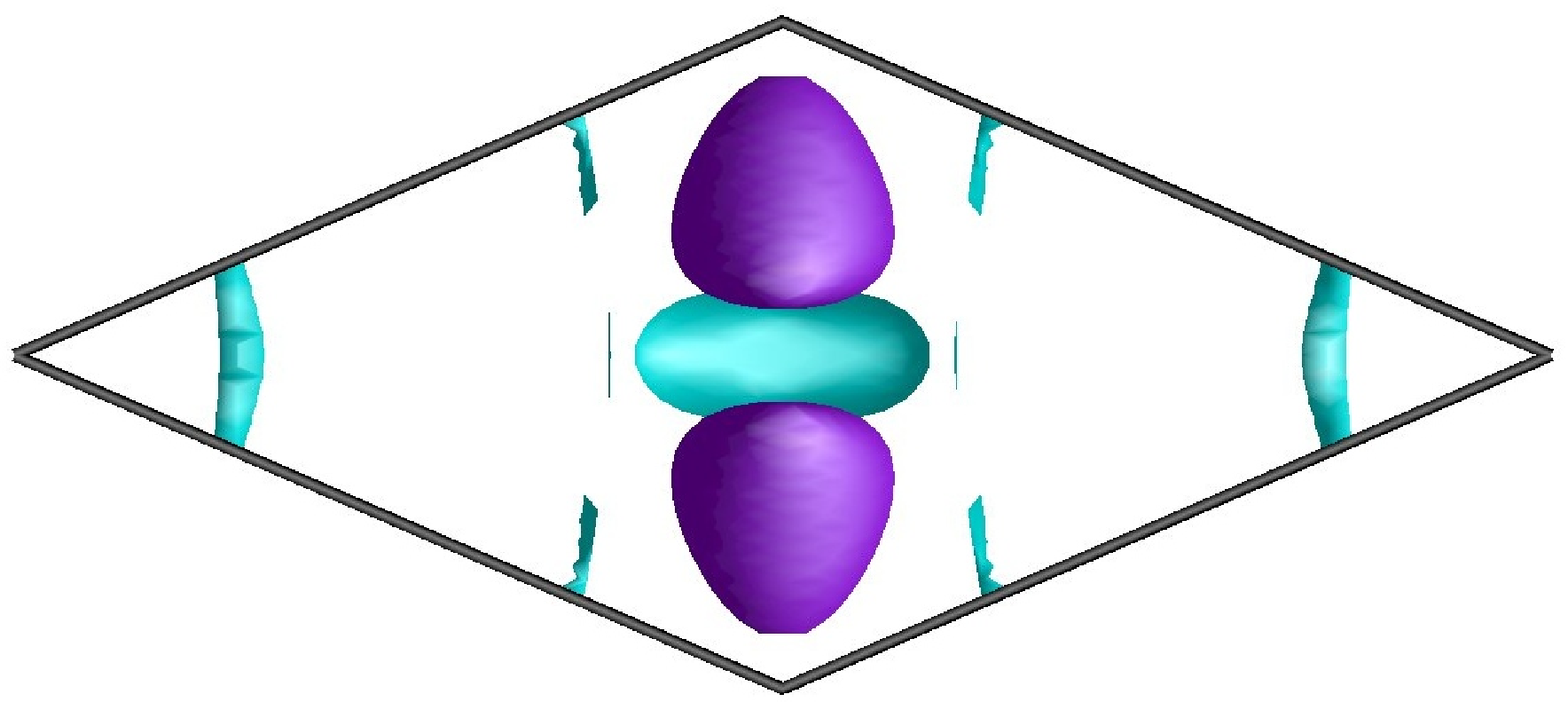}
\vspace{0.1cm}
\includegraphics[height=3.1cm,width = 4 cm]{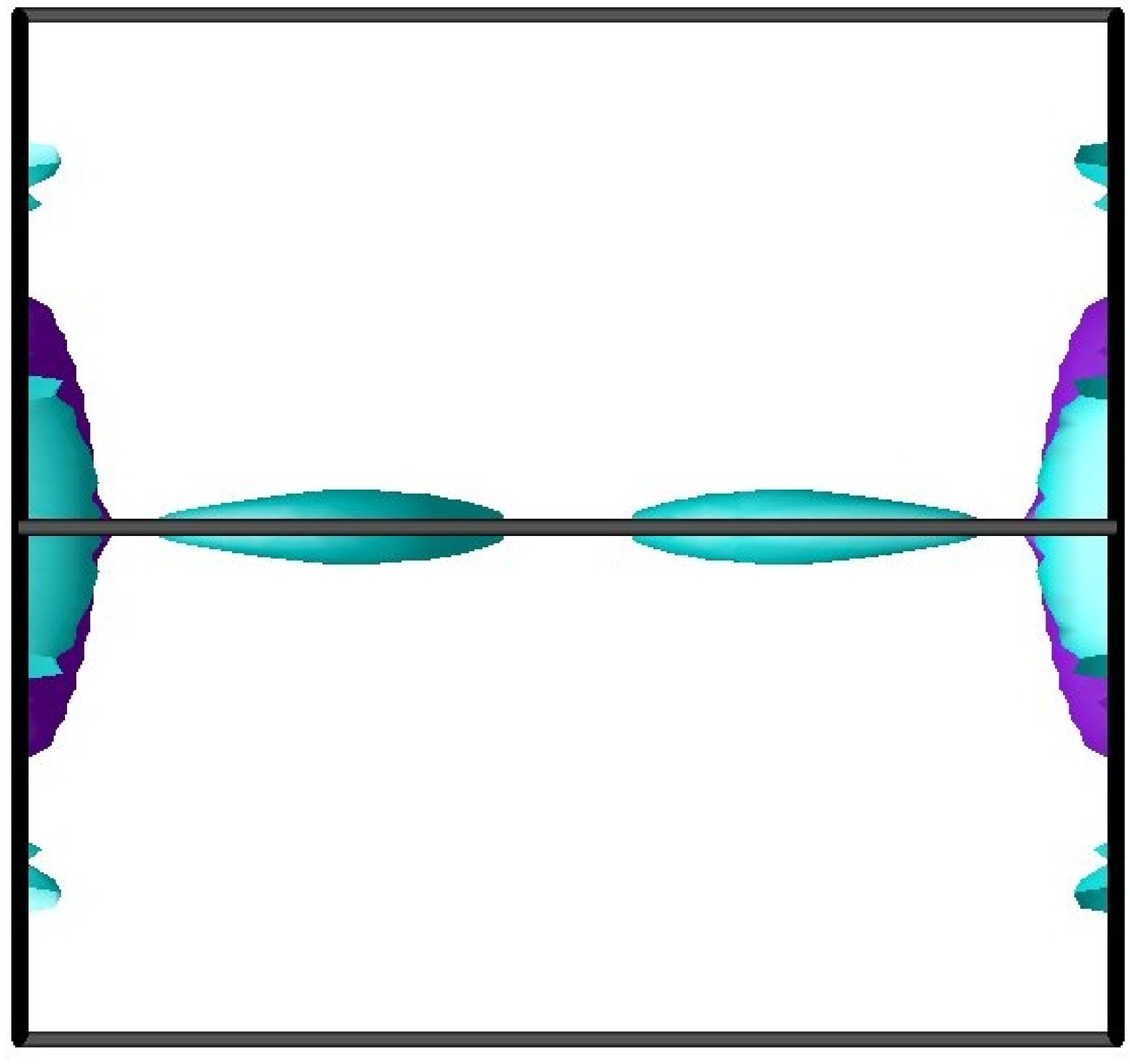}
\hspace{0.05cm}
\includegraphics[height=3.2cm,width = 4 cm]{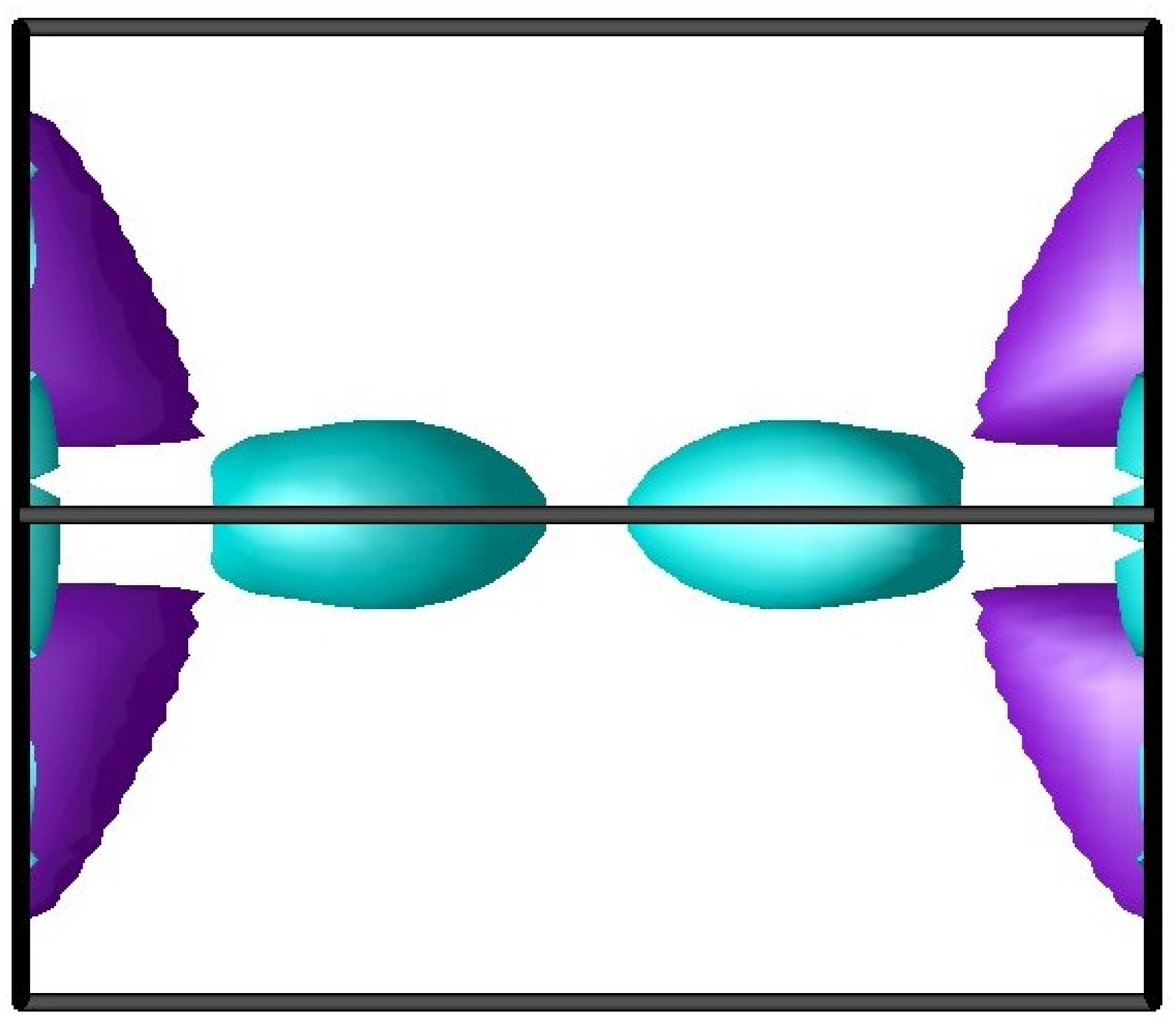}
\caption{(color online)
Fermi surfaces for Ca122 (left panels) and Ba122
(right panels), obtained in the LDA band structure calculation
(low magnetic moment). One reciprocal lattice unit cell is shown, with
the $\Gamma$ points located in the eight corners.  The top panels represent
a general view (directions of the axes as shown), the
next row represents the view from the top, along the $k_z$ direction  ($k_{x}$-$k_{y}$ plane),
the third row represents the front view, along the $k_y$ direction
($k_{x}$-$k_{z}$ plane), the last row presents the side view, along the $x$ direction
 ($k_{y}$-$k_{z}$ plane). The light shaded (cyan) surfaces are hole
pockets, the dark (blue) one are the electron pockets.
 }%
\label{FS}%
\end{figure}

Let us now analyze the most anisotropic term in Eq. \ref{rho}, the plasma
frequencies. These are intimately related with the electronic structure near
the Fermi level, so we show in Fig.~\ref{FS} the calculated Fermi surfaces for
both compounds in their calculated ground states. Note that while the
calculated ground state moment is \textit{larger} than the experimental, the
calculated number of carriers (the Fermi surface volume) is also larger that
the experimental one \cite{Mazin,alloul}. Therefore it is hard to say what
would be a better calculation to use for the transport properties: GGA gives
more correct (but still too large) volumes of the Fermi surfaces, while LDA
gives more correct (but still too large) magnetic moments. Besides, as we will
show next, the calculated anisotropy is very sensitive to these details,
primarily because of the small size and the complicated shape of the Fermi
surfaces. That is to say, the computational results should be taken with a
large grain of salt, and even if they provide in some aspects reasonable
quantitative agreement with the experiment, this is likely fortuitous and one
should not be tempted to use the calculated numbers any more than a
qualitative guide.

Nevertheless, one can readily make some qualitative observations: first, both
electron and hole Fermi surfaces strongly violate the tetragonal symmetry;
second, the two compounds differ substantially (this has been noticed before
\cite{Yildirim} and ascribed to the direct As-As overlap across the Ca plane).
One can naively think that the calculated in-plane transport anisotropy should
be very large, yet it appears not to be the case (Table 1). Particularly, in
the antiferromagnetic LDA ground state Ca122 has an anisotropy $\sigma
_{xx}/\sigma_{yy}$ of 1.10 ($+10\%),$ and in the GGA ground state 0.97
($-3\%).$ In Ba122 the corresponding numbers are $+14\%$ and $+31\%$.

\begin{table}[ptb]
\caption{ Calculated total and partial (hole, electrons) squared plasma
frequencies in eV$^{2}$. The $x$ axis is parallel to the crystallographic
$a_{o}$ axis and perpendicular to the ferromagnetic chain direction. The second column
indicates the calculated magnetic moment in $\mu_B$/Fe. }%
\label{table}
\begin{tabular}
[c]{l|lllllll}
&$M$& $\omega_{px}^{2}$ & $\omega_{py}^{2}$ & $\omega_{px}^{2}(h)$ & $\omega
_{py}^{2}(h)$ & $\omega_{px}^{2}(e)$ & $\omega_{py}^{2}(e)$\\\hline
Ca122 (GGA)& 1.85 & 0.36 & 0.37 & 0.14 & 0.24 & 0.22 & 0.14\\
Ca122 (LDA)& 1.61 & 0.46 & 0.42 & 0.11 & 0.24 & 0.35 & 0.18\\
Ba122 (GGA)& 1.96 & 0.42 & 0.32 & 0.13 & 0.13 & 0.29 & 0.19\\
Ba122 (LDA)& 1.64 & 0.41 & 0.36 & 0.20 & 0.17 & 0.20 & 0.20
\end{tabular}
\end{table}

Interestingly, the reason for such low anisotropy is different for the two
compounds. As one can see from the Table 1, in Ca122 the holes and the
electrons separately have noticeable anisotropies \textit{of the opposite
sign}: $-42\%$ and $+57\%,$ respectively, in GGA, and $-46\%$ and $+94\%$ in
LDA. The net anisotropy is thus small owing to the large cancellation of these
two terms. This cancellation holds only if the relaxation times for holes and
electrons are similar. The validity of this approximation in Ca122 is
unknown, but in Ba122 it is known to be highly questionable. Based on the Hall
effect data, it was argued \cite{Mazin,alloul} that  the
electrical transport in Ba122 at low temperatures is dominated by electrons.

On the other hand, in Ba122 both the holes and electrons separately show
rather weak anisotropy \textit{of the same sign, }especially in LDA: $18\%$
for the holes and $0\%$ for the electrons. In GGA, these numbers are $0$ and
$+52\%$. The reason is that while the Fermi surface pockets in Ba122 are
aligned along $x$, violating the tetragonal symmetry, each of these pockets
separately (and as opposed to Ca122) shows relatively little anisotropy.

While these numbers are not quantitatively reliable, they show two qualitative
trends observed in the experiment: firstly, despite the fact that the Fermi
surfaces themselves are strongly anisotropic, the transport anisotropy is weak
in Ca122 and moderate in Ba122. Secondly, despite the fact that the magnetic
structure suggests much higher conductivity along $y$ ($i.e,$along the
ferromagnetic chains) than along $x,$ the opposite holds both in the
calculations and in the experiment.

The temperature dependence is more intriguing. Obviously, if one could plot
the anisotropy as a function of the long-range staggered magnetization, in has to
be finite when the magnetization is maximal, at $T=0,$ and zero when
the magnetization is vanishingly small. Yet in the experiment the anisotropy
is growing with temperature, while the moment decreases. This means that
right below $T_{\mathrm{TO}}$ the magnetic moments are already large, even
though incompletely ordered (as oppose to what the fully spin-Peierls model
\cite{Chubukov, Eremin} would imply) and so is the transport anisotropy. An
example of such model is the dynamic domain scenario \cite{Johannes} that
suggests that the long-range order is destroyed primarily by fluctuating
domain walls that affect the average long-range magnetization, but do not
affect electronic anisotropy (see Fig.~1a in Ref.~\onlinecite{Johannes}).

Within this scenario we can understand why the anisotropy experiences a nearly
finite jump below $T_{\mathrm{TO}},$ however, one needs to explain why the
anisotropy actually decreases with cooling. One possible explanation is that
the ratio of the plasma frequencies nonmonotonically depends on the magnitude
of the magnetic moment. In this case anisotropy will be correlated with the
moment for very small (experimentally inaccessible) magnetization, but
anticorrelated with the moment at larger moments.

Alternatively, one may think that if the hole and the electron anisotropies
have opposite signs, the net anisotropy is defined by their incomplete
cancelation. In this case the observed temperature dependence is defined by
the temperature dependence of the $\tau_{h}/\tau_{e}$ ratio, which, of
course,
can have virtually any temperature dependence. This latter scenario is
corroborated by the calculations in Ca122, but not in Ba122. One has to
remember, as we mentioned before, that these calculations are not to be trusted
on a quantitative level.

\section{Conclusions}

In conclusion, we performed reversible mechanical detwinning of single crystal
CaFe$_{2}$As$_{2}$ and BaFe$_{2}$As$_{2}$. The single domain state was
confirmed by direct polarized light imaging in both compounds and with
high-energy $x$-rays in BaFe$_{2}$As$_{2}$. In both materials we find that the
resistive anisotropy is largest at $T_{\mathrm{TO}}$ with $\rho_{ao}/\rho
_{bo}\approx1.2$ in Ca122 and $\approx$1.5 in Ba122. For Ca122 this anisotropy
only exists below $T_{\mathrm{TO}}$ and diminishes upon further cooling,
reaching about 1.05 at $T\sim50$ K and remaining around that value at lower
temperatures. For Ba122 the anisotropy exists both below and above
$T_{\mathrm{TO}}$, reflecting nematic fluctuations above $T_{\mathrm{TO}}$ and
the second-order character of the phase transition there. The temperature
dependence of the anisotropy is weaker in Ba122, the low-temperature value
being comparable with the high-temperature one, $1.4$ vs. 1.5.

These results are rather counterintuitive in several aspects. First, if one
adapts a simple Jahn-Teller-like orbital ordering picture \cite{Phillip-Phillips,
Rajiv-Singh,Zaanen,Spain}, one expects that conductivity along the ferromagnetic
\textquotedblleft metallic\textquotedblright\ chains ($y)$ should be higher
than along the antiferromagnetic \textquotedblleft insulating\textquotedblright%
\ direction ($x).$ The opposite is true. Second, given the drastic anisotropy
of the band structure and the Fermi surfaces (including the fully broken
symmetry between the $xz$ and $yz$ orbitals), as confirmed by both first
principles calculations and the Hall experiments, one expects a dramatic
anisotropy, while in fact we observe weak to moderate anisotropy, not atypical
for many metals. Third, within the experimental accuracy, the maximal
anisotropy is right below the transition, and not when the order parameter
(magnetic moment or the orthorhombicity) is maximal. Finally, despite the
fact that the degree of orthorhombicity and the magnetic order parameter
monotonically $decrease$ with temperature, the transport anisotropy actually
$increases.$

The first two findings are fully corroborated by the first principle
calculations. These predict the correct (counterintuitive) sign of the
anisotropy, and a relatively moderate anisotropy magnitude. The third might reflect the
existence of dynamic domains (similar to the nematic fluctuations) right below
$T_{\mathrm{TO}}$ in Ca122, and both below and above in Ba122. In that case,
the antiphase domain walls, as described in \cite{Johannes}, break the long
range order without destroying the transport anisotropy.

As to the temperature dependence, we have no preferable interpretation of this
effect; the most plausible cause seems to be the temperature dependence of the
relaxation rates, which are known to change by more than an order of magnitude
between $T=0$ and $T=T_{\mathrm{TO}}$.

While this manuscript was prepared for submission, a preprint appeared
reporting a partial mechanical detwinning using uniaxial stress in
Ba(Fe$_{1-x}$Co$_{x}$)$_{2}$As$_{2}$ \cite{fisher2}. The reported resistivity data for
the parent BaFe$_{2}$As$_{2}$ are similar to those found in our study.

We thank D. Robinson for the excellent support of the high--energy x-ray scattering study and L. Podervyanski for help in writing the manuscript. Use of the Advanced Photon Source was supported by the U. S. Department of Energy, Office of Science, under Contract No. DE-AC02-06CH11357. Work at the Ames Laboratory was supported by the U.S. Department of Energy, Office of Basic Energy Sciences, Division of Materials Sciences and Engineering under contract No. DE-AC02-07CH11358.  R. P. acknowledges support from Alfred P. Sloan Foundation.



\begin{thebibliography}{99}
\bibitem {c-axis}The $x$-ray study of Ref.~\onlinecite{domains} revealed that
despite small size, the domains propagate through the whole sample thickness
and are of columnar type.
\end{thebibliography}
\end{document}